\documentclass{llncs}
\usepackage{graphicx}
\usepackage{url}

\usepackage{hyperref}

\title{Parallelizing Mizar}
\author{Josef Urban\thanks{Supported by the NWO
project MathWiki.}}
\institute{Radboud University, Nijmegen}

\makeatletter
\renewcommand\section{\@startsection{section}{1}{\z@}%
                       {-12\p@ \@plus -4\p@ \@minus -4\p@}%
                       {8\p@ \@plus 4\p@ \@minus 4\p@}%
                       {\normalfont\large\bfseries\boldmath
                        \rightskip=\z@ \@plus 8em\pretolerance=10000 }}
\makeatother

\begin{document}
\maketitle

\begin{abstract}
  This paper surveys and describes the implementation of
  parallelization of the Mizar proof checking and of related Mizar
  utilities. The implementation makes use of Mizar's compiler-like
  division into several relatively independent passes, with typically
  quite different processing speeds. The information produced in
  earlier (typically much faster) passes can be used to parallelize
  the later (typically much slower) passes. The parallelization now
  works by splitting the formalization into a suitable number of
  pieces that are processed in parallel, assembling from them together
  the required results. The implementation is evaluated on examples
  from the Mizar library, and future extensions are discussed.
\end{abstract}
%------------------------------------------------------------------------------

\section{Introduction and Motivation}
While in the 90-ies the processing speed of a single CPU has grown
quickly, in the last decade this growth has considerably slowed down,
or even stopped.  The main advances in processing power of computers
have been recently done by packing multiple cores into a single CPU,
and related technologies like hyperthreading.  A low-range dual-CPU
(Intel Xeon 2.27 GHz) MathWiki server of the Foundations Group at the
Radboud University bought in 2010 has eight hyperthreading cores, so
the highest raw performance is obtained by running sixteen processes
in parallel. The server of the Mizar group at University of Bialystok
has similar characteristics, and the Mizar server at University of
Alberta has twelve hyperthreading cores.  
%Intel's Westmere-EX 10-core
%processor will be shipped in the first half of 2011, available in
%eight-socket configurations. 
%With each physical core being able to run
%two threads, such servers will have the capability to run 160 threads
%simultaneously.  
Packing of CPU cores together is happenning not only
on servers, but increasingly also on desktops and notebooks, making
the advantages of parallelization attractive to many applications.

To take advantage of this development, reasonable ways of
parallelizing time-consuming computer tasks have to be
introduced. This paper discusses the various ways of parallelization
of proof checking with the Mizar formal proof verifier, and
parallelization of the related Mizar utilities.  Several
parallelization methods suitable for different scenarios and use-cases
are introduced, implemented, and evaluated.

The paper is organized as follows: Section~\ref{MizarProcessing}
describes the main tasks done today by the
Mizar~\cite{mizar-in-a-nutshell,RudnickiT99} verifier and related utilities,
and the ways how they are performed.  Section~\ref{PossibleWays}
explores the various possible ways and granularity levels in which
suitable parallelization of the Mizar processing could be done, and
their advantages and disadvantages for various use scenarious.
% Makefile parallelization useful for mathwiki
Section~\ref{Makefile} describes and evaluates 
parallelization of the processing of the whole Mizar library and Mizar wiki
done on the coarsest level of granularity, i.e. on the article level.
Section~\ref{Mizp} then describes the recent parallelization done on
sub-article levels of granularity, i.e. useful for the speedup of
processing of a single Mizar article. Both the verification and
various other utilities have been parallelized this way, and
evaluation on hundreds of Mizar articles is done.  Section~\ref{Future}
names possible future directions, and concludes.

\section{Mizar Processing}
\label{MizarProcessing}
\subsection{Article Workflow}
The term \emph{Mizar Processing} can in the broad sense refer to
several things.  Mizar consists of a large library of formal
matheamatical articles, on top of which new articles are written,
formally verified by the Mizar verifier, possibly also checked by
various (proof improving) utilities during or after the writing,
possibly HTML-ized for better understanding during and after the
writing, and usually translated to TeX after they are written. During
the verification a number of tools can be used, ranging from tools for
library searching, tools for creating proof skeletons, to tools for
ATP or AI based proof advice.

After a new article is written, it is typically submitted to the
library, possibly causing some refactoring of the library and itself,
and the whole new version of the library is re-verified (sometimes
many times during the refactoring process), and again possibly some
more utilities can be then applied (again typically requiring further
re-verification) before the library reaches the final state. The new
library is then HTML-ized and publicly released. The library also lives
in the experimental Mizar wiki based on the git distributed version
control system~\cite{UrbanARG10,AlamaBMU11}. There, collaborative re-factoring of
the whole library is the main goal, requiring fast real-time
re-verification and HTML linking.

\subsection{Basic Mizar Verification}
\label{Basic}
In more detail, the basic verification of an article starts by
selecting the necessary items from the library (so called
\emph{accommodation}) and creating an article-specific local
environment (set of files) in which the article is then verified
without further need to access the large library.  The verification
and other Mizar utilities then proceeds in several compiler-like
passes that typically vary quite a lot in their processing times. The
first \emph{Parser} pass tokenizes the article and does a fast
syntactic analysis of the symbols and a rough recognition of the main
structures (proof blocks, formulas, etc.).

The second \emph{Analyzer} pass then does the complete type
computation and disambiguation of the overloading for terms and
formulas, and checks the structural correctness of the natural
deduction steps, and computes new goals after each such step. These
processes typically take much longer than the parsing stage,
especially when a relatively large portion of the library is used by
the article, containing a large amount of type automations and
overloaded constructs. The main product of this pass is a detailed XML
file containing the disambiguated form of the article with a number of
added semantic information~\cite{Urban05}.  This file serves as the main input for
the final \emph{Checker} pass, and also for the number of other Mizar
proof improving utilities (e.g., the
\emph{Relprem}\footnote{Irrelevant Premises Detector} utility
mentioned in Table~\ref{Speeds}), for the HTML-ization, and also for
the various ATP and AI based proof advice tools.

The final \emph{Checker} pass takes as its main input the XML file
with the fully disambiguated constructs, and uses them to run the
limited Mizar refutational theorem prover for each of the (typically
many) atomic (\emph{by}) justification steps. Even though this checker
is continuosly optimised to provide a reasonable combination of
strength, speed, and ``human obviousness'', this is typically the
slowest of the verifier passes. Similar situation is with the various
utilities for improving (already correct) Mizar proofs.  Such
utilities also typically start with the disambiguated XML file as an
input, and typically try to merge some of the atomic proof steps or
remove some redundant assumptions from them. This may involve running
the limited Mizar theorem prover several times for each of the atomic
proof steps, making such utilities even slower than the \emph{Checker}
pass.

\subsection{Other Tools}
All the processes described so far are implemented using the Mizar
code base written in object-oriented extension of
Pascal. % This (thanks to the existence of
The disambiguated XML file is also used as an input for creation of
the HTML representation of the article, done purely by XSL
processing. XSL processing is also used for translation of the article
to an ATP format, serving as an input for preparing ATP problems
(solvable by ATP systems) corresponding to the problems in the Mizar
article, and also for preparing data for other proof advice systems
(MML Query, Mizar Proof Advisor). The XSL processing is usually done
in two stages. The first stage (called
\emph{absolutization}) is common for all these utilities, it basically
translates the disambiguated constructs living in the local article's
environment into the global world of the whole Mizar library. The
second stage is then the actual XSL translation done for a particular
application. The XSL processing can take very different times
depending on its complexity. Generally, XSL processors are not as much
speed-optimized as, e.g., the Pascal compilers, so complex XSL
processing can take more time than analogous processing programmed in
Pascal.

Finally, there are a number of proof advice tools, typically taking as
input the suitably translated XML file, and providing all kinds of
proof advice using external processing. Let us mention at least the
Automated Reasoning for Mizar~\cite{UrbanS10,abs-1109-0616,KaliszykU13b} system, linking Mizar
through its Emacs authoring environment and through a HTML interface
to ATP systems (particulary a custom version~\cite{UrbanHV10} of the
Vampire-SInE system~\cite{RiazanovV02} and a customized~\cite{blistr} version of E~\cite{Sch02-AICOMM}) usable for finding and
completing proofs automatically, for explaining the Mizar atomic
justifications, and for ATP-based cross-verification of Mizar.  This
processing adds (at least) two more stages: (i) It uses the MPTP
system~\cite{Urb04-MPTP0,Urban06-jar} to produce the ATP problems corresponding to
the Mizar formulation, and (ii) it uses various ATP/AI systems and
metasystems to solve such problems. Attached to such functions is
typically various pre/post-processing done in Emacs Lisp and/or as CGI
functions.

\begin{figure}[htbp]
  \caption{Structure of the Mizar processing for one article}
    \includegraphics[height=95mm,width=110mm]{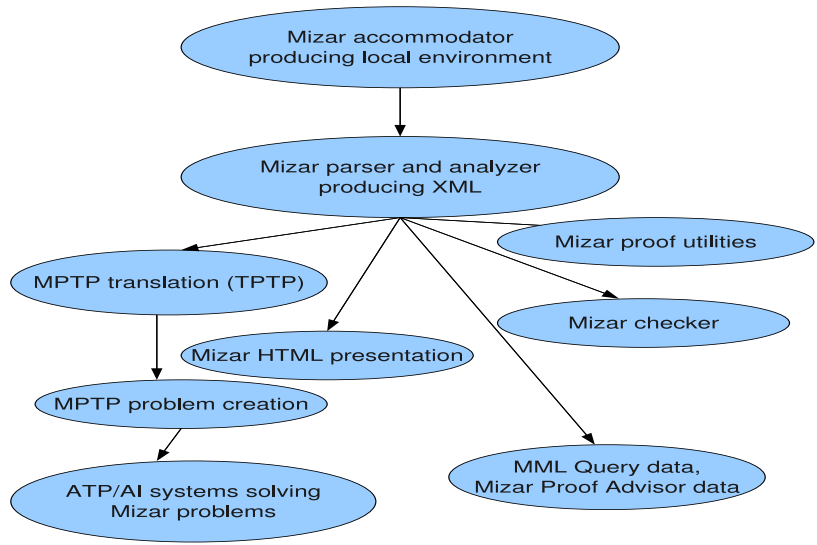}
  \label{Structure}
\end{figure}

See Figure~\ref{Structure} for the overall structure of Mizar and
related processing for one article. Table~\ref{Speeds} gives timings
of the various parts of Mizar processing for the more involved Mizar
article {\tt fdiff\_1} about real function
differentiability\footnote{%
  \url{http://mws.cs.ru.nl/~mptp/mml/mml/fdiff_1.miz}} \cite{FDIFF1},
and for the less involved Mizar article {\tt abian} about Abian's
fixed point theorem\footnote{%
  \url{http://mws.cs.ru.nl/~mptp/mml/mml/abian.miz}} \cite{ABIAN} run
on recent Intel Atom 1.66 GHz notebook\footnote{This small measurement
  is intentionally done on a standard low-end notebook, while the rest
  of global measurements in this paper are done on the above mentioned
  server of the Foundations Group. This is in order to compare the
  effect of parallelized server-based verification with standard notebook work in
  Section~\ref{Mizp}.}.
\vspace{-5mm}
\vspace{-1cm}
\begin{table*}[hbp]
  \caption{Speed of various parts of the Mizar processing on articles fdiff\_1 and abian in seconds - real time and user time}
\begin{center}
  \begin{tabular}{|l|r|r|r|r|}
    \hline
    Processing (language)&real - fdiff\_1 &user - fdiff\_1&real - abian&user - abian\\
    \hline
    Accommodation (Pascal)&1.800&1.597&1.291&1.100\\
    \hline    
    Parser (Pascal)&0.396&0.337&0.244&0.183\\
    \hline    
    Analyzer (Pascal)&28.455&26.155&4.182&4.076\\
    \hline    
    Checker (Pascal)&39.213&36.631&10.628&10.543\\
    \hline    
    Relprem (Pascal)&101.947&99.385&48.493&47.683\\
    \hline    
    Absolutizer (XSL)&17.203&13.579&9.624&7.886\\
    \hline    
    HTML-izer (XSL)&27.699&24.498&11.582&11.323\\
    \hline    
    MPTP-izer (XSL)&70.153&68.919&47.271&45.410\\
    \hline    
  \end{tabular}
\end{center}
\label{Speeds}
\end{table*}

% graph of the processes
%cost table done on mws for a couple of articles

\section{Survey of Mizar Parallelization Possibilities}
\label{PossibleWays}
There are several ways how to parallelize Mizar and related utilities,
and several possible levels of granularity. Note that for any of these
Mizar parallelization methods the main issue is speed, not the memory
consumption. This is because Pascal does not have garbage collection,
and Mizar is very memory efficient, taking typically less than 30MB
RAM for verifying an article. The reason for this extreme care is
mainly historical, i.e., the codebase goes back to times when memory
was very expensive. Methods used for this range from exhaustive
sharing of data structures, to using only the part of the library that
is really necessary (see \textit{accommodation} in~\ref{Basic}).

The simplest method of parallelization which is useful for the Mizar
wiki users, developers, and library maintainers is article-level
parallelization of the whole library verification, and parallization
of various other utilities applied to the whole Mizar library. There
are about 1100 Mizar articles in the recent library, and with this
number the parallelization on the article level is already very useful
and can bring a lot of speed-ups, especially useful in the real-time
wiki setting, and for the more time consuming utilities like the above
mentioned \emph{Relprem}.

A typical user is however mainly interested in working with one (his
own) article. For that, finer (sub-article) levels of parallelization
are needed. A closer look at the Table~\ref{Speeds} indicates that the
\emph{Parser} pass of the verification is very fast, while the
\emph{Analyzer} and especially the \emph{Checker} passes are the
bottlenecks (see also the global statistics for the whole MML
processing done with article-level parallelization in
Table~\ref{MMLSpeeds}).
% MPTP parallelism, ATP parallelism (different timelimits, malarea), calling ATPs parallelism

\subsection{Checker parallelization}
There are several basic options to parallelizing the most costly 
verification operation - the \emph{Checker} pass, they are explained in more detail below:
\begin{enumerate}
\item Running several \emph{Checker} passes in parallel as separate
  executables, each checking only a part of the atomic steps conducted
  in the article
\item Running one \emph{Checker} pass as only one executable, with
  multithreading code used for parallelizing the main checking
  procedure
\item Running one \emph{Checker} pass as only one executable, with
  multithreading code used inside the main checking procedure
\item Combinations of above
\end{enumerate}

As mentioned above, the input for the \emph{Checker} pass is a fully
disambiguated article, where only the atomic justification steps need
to be checked, i.e. proved by the Mizar's limited theorem prover. The
number of such atomic justification steps in one article is typically
high, about every second to third line in a formal Mizar text is
justified in such a way.  The result of one such theorem proving attempt
is completely independent of others, and it is just a boolean value
(true or false)\footnote{Note that this is not generally true for
  nonclassical systems like Coq, where the proof might not be an
  opaque object.}. All of these theorem proving attempts however share
a lot of data-structures that are basically read-only for them, for
example information about the types of all the ground terms appearing
up to the particular point in the formal text, and information about
the equalities holding about ground terms at particular points of the
formal text.

The first method suggested above - running several \emph{Checker}
passes in parallel as separate executables, each checking only a part
of the atomic steps conducted in the article - is relatively
``low-tech'', however it has some good properties. First, in the
methods based on multithreading, the relatively large amount of the
shared data has to be cloned in memory each time a new thread is
created for a new justification step. This is not the case when
several executables are running from the beginning to the end, each
with its own memory space. Second, the implementation can be
relatively simple, and does not require use of any multithreading
libraries, and related refactoring of the existing single-threaded
code.

The second and third method require the use of a multithreading
library (this is possible for the Free Pascal Compiler used for Mizar,
with the \emph{MTProcs} unit), and related code refactoring. There are
several places where the multithreading can be introduced relatively
easily, let us name at least the most obvious two: (i) the main entry
to the refutational proof checker, and (ii) within the refutational
proof checker, separately disproving each of the disjuncts in the
toplevel disjunctive normal form created in the initial normalization
phase. The advantage of such implementation in comparison with running
several executables would probably be more balanced load, and in the
latter case, possibly being able to use more extreme parallelization
possibilities (e.g., if 1000 cores are available, but the article has
only 500 atomic justifications).

\subsection{Type Analysis and Caching: Why not use fine
  multithreading}
\label{Types}
\subsubsection{Caching vs. Multithreading}
For the also relatively costly \emph{Analyzer} pass, the methods based
on fine multithreading however seem to be either relatively
complicated or of relatively little value. The problem is following: A
major and increasing amount of work done in \emph{Analyzer} consists
in computing the full types of terms. This is because the Mizar
mechanisms for working with adjectives are being used more and more,
and are being made stronger and stronger, recently to a level that
could be compared to having arbitrary Prolog programs working over a
finite domain (a finite set of ground terms). The method that then
very considerably improves the \emph{Analyzer} efficiency in the
singlethreaded case is simple caching of terms' types. With a simple
multithreaded implementation, when the newly computed types are
forgotten once the thread computing them exits, this large caching
advantage is practically lost. Implementation where each thread
updates the commonly used cache of terms' types is probably possible,
but significantly more involved, because the access to the shared
datastructures is then not just read-only (like in the \emph{Checker}
case), and the updates are likely to be very frequent.

\subsubsection{Suitable Parallelization for Tree-like Documents}
Above is the reason why in the \emph{Analyzer} case, it makes much more
sense to rather have several ``long-term-running'' threads or
processes, each developing and remembering its own cache of terms'
types. The main problem is then to determine a proper level of
granularity for dividing \emph{Analyzer}'s work into such larger
parts. Unlike in the \emph{Checker} pass, \emph{Analyzer} is not a
large set of independent theorem proving runs returning just a boolean
result. Analysing each term depends on the analysis of its subterms,
and similarly, analysing the natural deduction structure of the proofs
(another main task of this pass) depends on the results of the
analysis of the proof's components (formulas, and natural deduction
steps and subproofs).  Thus, the finer the blocks used for
parallelization, the larger the part that needs to be repeated by
several threads (all of them having to analyse all the necessary parts
of the nested proof, formula, and term levels leading to the fine
parallelized part). To put this more visually, the formal text (proof,
theory) is basically a tree (or forest) of various dependencies.  The
closer to the leaves the parallelization happens, the more common work
has to be repeated by multiple threads or processes when descending
down the branches to the parallelization points on those
branches. Obviously, the best solution is then to parallelize not on
the finest possible level, but on the coarsest possible level, i.e.,
as soon as there are enough branches for the parallelization.

\subsubsection{Toplevel Proofs as Suitable Parallelization Entry Points}
To this requirement reasonably corresponds the choice of toplevel
proofs in a given formal text as the entry points for
parallelization. There are typically tens to hundreds of toplevel
proofs in one article, and with some exceptions (very short articles,
or articles consisting of one very involved proof) these toplevel
proofs can usually be divided into the necessary number of groups with
roughly the same overall length. Mizar (unlike e.g. Coq) never needs
the proofs for anything, only the proved theorem can be used in later
proofs. Thanks to this, a simple directive ({\tt @proof}) was
introduced in the Mizar language long time ago, in order to omit
verification of the (possibly long) proofs that have already been
proved, and would only slow-down the verification of the current
proof. This directive basically tells to the \emph{Parser} to skip all
text until the end of the proof is found, only asserting the
particular proposition proved by this skipped proof. Due to the
file-based communication between the passes, the whole skipped proof
therefore never appears in the \emph{Analyzer}'s input, and
consequently is never analyzed. This feature can be used for 
file-based parallelization of the \emph{Analyzer}, described in more
detail in Section~\ref{Mizp}. It also parallelizes the \emph{Checker},
and also can be used for easy parallelization of the subsequent
HTML-ization.

\subsection{HTML-ization parallelization}
\label{HTML}
As mentioned above, HTML-ization of Mizar texts is based on the
disambiguated article described in the XML file produced by the
\emph{Analyzer}. HTML-ization is done completely separately from the
Mizar codebase written in Pascal, by XSL processing. Even though XSL
is a pure lazily evaluated functional language\footnote{Thanks to
  being implemented in all major browsers, XSL is today probably by
  far the most widely used and spread purely functional language.}, as
of January 2011, the author is not aware of a XSL processor
implementing multithreading. The remaining choice is then again
file-based parallelization, which actually corresponds nicely to the
file-based parallelization usable for skipping whole proof blocks in
the \emph{Analyzer}. During the XSL processing, it is easy to put the
HTML-ized toplevel proofs each into a separate file\footnote{This
  functionality actually already exists independently for some time,
  in order to decrease the size of the HTML code loaded into browser,
  loading the toplevel proofs from the separate files by AJAX calls.},
and then either to load the proofs into a browser on-demand by AJAX
calls, or to merge the separate files with HTML-ized proofs created by
the parallelization by a simple postprocessing into one big HTML file.

\subsection{Parallelization of Related Mizar Processing}
Remaining Mizar refactoring utilities (like \emph{Relprem}) are
typically implemented by modifying or extending the \emph{Checker} or
\emph{Analyzer} passes, and thus the above discussion and solutions
apply to them too. Creation of data for MML Query, Mizar Proof
Advisor, and similar systems is done purely by XSL, and the file-based
approach can again be applied analogously to HTML-ization. The same
holds for translating the article to the MPTP format (extended TPTP),
again done completely in XSL. A relatively important part used for the
automated reasoning functions available for Mizar is the generation of
ATP problems corresponding to the Mizar problems.
% cite LPAR
This is done by the MPTP system implemented in Prolog. The problem
generating code is probably quite easily parallelizable in
multithreaded Prologs (Prolog is by design one of the most simply
parallelizable languages), however the easiest way is again just to
run several instances of MPTP in parallel, each instructed to create
just a part of all the article's ATP problems. The recent Emacs
authoring interface for Mizar implements the functions for
communicating with ATP servers asynchronously~\cite{RudnickiU11}, thus allowing to solve as many
ATP-translated problems in parallel as the user wants (and the
possible remote MPTP/ATP server allows). The asynchronously provided
ATP solutions then (in parallel with other editing operations) update
the authored article using Emacs Lisp callbacks.\footnote{See, e.g.,the
  AMS 2011 system demonstration at
  \url{http://mws.cs.ru.nl/~urban/ams11/out4.ogv}}

As for the parallelization of the ATP solving of Mizar problems, this
is a field where a lot of previous research
exists~\cite{SS99-FLAIRS,Sut01-LPAR}, and in some systems
(e.g. Waldmeister, recent versions of Vampire used for the Mizar ATP
service) this functionality is readily available. Other options
include running several instances of the ATPs with different
strategies, different numbers of most relevant axioms, etc.  The
MaLARea~\cite{Urban07-esarlt,UrbanSPV08,malar14} metasystem for solving
problems in large Mizar-like theories explores this number of choices
in a controlled way, and it already has some parallelization options
implemented.

\section{Parallelization of the MML Processing on the Article Level}
\label{Makefile}
% The historically first parallelization of Mizar processing was done in
% early 2006, when the author acquired a four-core AMD server, and
% wanted to use it for conducting the newly implemented HTMLization of
% the whole Mizar library. Later the need to regularly translate the
% whole library into the MPTP format arose, together with the need to
% produce a large number of ATP problems corresponding to Mizar, and the
% need to solve as many of them by ATPs in as low time as possible by
% systems like MaLARea. The benefits from potential parallelization
% further improved by acquiring an eight-core Intel server in early
% 2008.

A strong motivation for fast processing of large parts of the library
comes with the need for collaborative refactoring.  As the library
grows, it seems that the number of submissions make it more and more
difficult for the small core team of the library maintainers and
developers to keep the library compact, and well organized and
integrated together. The solution that seems to work for Wikipedia is
to outsource the process of library maintanance and refactoring to a
large number of interested (or addicted) users, through a web
interface to the whole library. In order for this to work in the
formal case, it is however important to be able to quickly re-verify
the parts of the library dependent on the refactored articles, and
notify the users about the results, possibly re-generating the HTML
presentation, etc.

The implementation of article-level parallelization is as
follows. Instead of the old way of using shell (or equivalent MS
Windows tools) for processing the whole library one article after
another, a Makefile has been written, using the files produced by the
various verification passes and other tools as targets, possibly
introducing artificial (typically empty file) targets when there is no
clear target of a certain utility. The easiest option once the various
dependencies have been reasonably stated in the Makefile, is just to
use the internal parallelization implemented in the GNU \emph{make}
utility. This parallelization is capable of using a pre-specified
number of processes (via the {\tt -j} option), and to analyse the
Makefile dependencies so that the parallelization is only done when
the dependencies allow that. The Makefile now contains dependencies
for all the main processing parts mentioned above, and is regularly
used by the author to process the whole MML and generate HTML and data
for various other tools and utilities. In Table~\ref{MMLSpeeds} the
benefits of running {\tt make -j64} on the recently acquired
eight-core hyperthreading Intel Xeon 2.27 GHz server are
summarized. The whole library verification and HTML-ization process
that with the sequential processing can take half a day (or much more
on older hardware), can be done in less than an hour when using this
parallelization. See~\cite{UrbanARG10} for further details and challenges
related to using this technique in the git-based formal Mizar wiki backend to
provide reasonably fast-yet-verified library refactoring.
%mathwiki usefulness
%theretical limits of finer parallelizations
%graph of the article depd using dotty

\begin{table*}[htbp]
  \caption{Speed of various parts of the Mizar processing on the MML (1080 articles) with 64 process parallelization run on an 8-core  hyperthreading machine, in seconds - real time and user time, total and averages for the whole MML.}
\begin{center}
  \begin{tabular}{|l|r|r|r|r|}
    \hline
    Stage (language) & real times total & user times total & real times avrg & user times avrg \\
    \hline    
    Parser (Pascal)&14 &91& 0.01 & 0.08\\
    \hline    
    Analyzer (Pascal)&330&4903& 0.30 & 4.53 \\
    \hline    
    Checker (Pascal)&1290 &18853& 1.19 & 17.46  \\
    \hline    
    Absolutizer (XSL)&368&4431& 0.34 & 4.10 \\
    \hline    
    HTML-izer (XSL)&700&8980& 0.65 & 8.31 \\
    \hline    
  \end{tabular}
\end{center}
\label{MMLSpeeds}
\end{table*}
\vspace{-8mm}
Similar Makefile-based parallelization technology is also used by the
MaLARea system when trying to solve the ca.  fifty thousand Mizar
theorem by ATPs, and producing a database of their solutions that is
used for subsequent better proof advice and improved ATP solving using
machine learning techniques.  One possible (and probably very useful)
extension for purposes of such fast real-time library re-verification
is to extract finer dependencies from the articles (e.g. how
theorems depend on other theorems and definitions - this is already to
a large extent done e.g. by the MPTP system), and further speed up
such re-verification by checking only certain parts of the dependent
articles, see~\cite{AlamaMU12} for detailed
analysis. This is actually also one of the motivations for the
parallelization done by splitting articles into independently verified
pieces, described in the next section.

\section{Parallelization of Single Article Processing}
\label{Mizp}
%emacs
While parallelization of the whole (or large part of) library
processing is useful, and as mentioned above it is likely going to
become even more used, the main use-case of Mizar processing is when a
user is authoring a single article, verifying it quite often. In the
case of a formal mathematical wiki, the corresponding use-case could
be a relatively limited refactoring of a single proof in a larger
article, without changing any of the exported items (theorems,
definitons, etc.), and thus not influencing any other proofs in any
other article. The need in both cases is then to (re-)verify the
article as quickly as possible, in the case of wiki also quickly
re-generating the HTML presentation, giving the user a real-time
experience and feedback.

\subsection{Toplevel Parallelization}
As described in Section~\ref{PossibleWays}, there are typically
several ways how to parallelize various parts of the processing,
however it is also explained there that the one which suits best the
\emph{Analyzer} and HTML-ization is a file-based parallelization over
the toplevel proofs. This is what was also used in the initial
implementation of the Mizar
parallelizer\footnote{\url{http://github.com/JUrban/MPTP2/raw/master/MizAR/cgi-bin/bin/mizp.pl}}. This
section describes this implementation (using Perl and LibXML) in more
detail.

As can be seen from Table~\ref{Speeds} and Table~\ref{MMLSpeeds}, the
\emph{Parser} pass is very fast. The total user time for the whole MML
in Table~\ref{MMLSpeeds} is 91.160 seconds, which means that the
average speed on a MML article is about 0.1 second. This pass
identifies the symbols and the keywords in the text, and the overall
block structure, and produces a file that is an input for the much
more expensive \emph{Analyzer} pass. Parsing a Mizar article by
external tools is (due to the intended closeness to mathematical
texts) very hard~\cite{CairnsG04}, so in order to easily identify the
necessary parts (toplevel proofs in our case) of the formal text, the
output of the \emph{Parser} pass is now also printed in an XML format,
already containing a lot of information about the proof structure and
particular proof positions\footnote{Note that the measurement of
  Parser speed in the above tables was done after the XMLization of
  the Parser pass, so the usual objection that printing a larger XML
  file slows down verification is (as usual) completely misguided,
  especially in the larger picture of costly operations done in the
  Analyzer and the Checker.}

The Parallelizer's processing therefore starts by this fast Parser
run, putting the necessary information in the XML file.  This XML file
is then (inside Perl) read by the LibXML functions, and the toplevel
proof positions are extracted by simple XPath queries from it. This is
also very fast, and adds very little overhead.  These proof positions
are an input to a (greedy) algorithm, which takes as another input
parameter the desired number of processes ($N$) run in parallel (for
compatibility with GNU make, also passed as the {\tt -j} option to the
parallelizer). This algorithm then tries to divide the toplevel proofs
into $N$ similarly hard groups. While there are various options how to
estimate the expected verification hardness of a proof, the simplest
and reasonably working one is the number of lines of the proof. Once
the toplevel proofs are divided into the $N$ groups, the parallelizer
calls Unix {\tt fork()} on itself with each proof group, spawning $N$
child instances.

Each instance creates its own subdirectory (symbolically linking there
the neccessary auxiliary files from the main directory), and creates
its own version of the verified article, by replacing the keyword {\tt
  proof} with the keyword {\tt @proof} for all toplevel proofs that do
not belong to the proofs processed by this particular child
instance. The \emph{Parser} pass is then repeated on such modified
input by the child instance, the {\tt @proof} directives producing
input for \emph{Analyzer} that contains only the desired toplevel
proofs. The costly subsequent passes like the \emph{Analyzer},
\emph{Checker},and HTML-ization can then be run by the child instance
on the modified input, effectively processing only the required
toplevel proofs, which results in large speedups. Note that the
Parser's work is to some extent repeated in the children, however its
work in the skipped proofs is very easy (just counting brackets that
open and close proofs), and this pass is in comparison with others very
fast and thus negligible. The parallel instances of the Analyzer,
Checker, and HTML-ization passes also overlap on the pieces of the
formal text that are not inside the toplevel proofs (typically the
stated theorems and definitions have to be at least analyzed), however
this is again usually just a negligible share of the formal text in
comparison with the full text with all proofs. 

The speedup measured for the verification (Parser, Analyzer, Checker)
passes on the above mentioned article {\tt fdiff\_1} run with eight parallel processes
{\tt -j8} is given in the Table~\ref{Par} below. While the total user
time obviously grows with the number of parallel processes used, the
real verification time is in this case decreased nearly four
times. Additionally, in comparison with the notebook processing
mentioned in the initial Table~\ref{Speeds}, the overall real-time
benefit of remote parallelized server processing is a speedup factor
of 20. This is a strong motivation for the server-based remote
verification (and other) services for Mizar implemented in Emacs and
through web interface decribed in~\cite{UrbanS10}.
The overall statistics done
across all (395) MML articles that take in the normal mode more than
ten seconds to verify is computed for parallelization with one, two,
four, and eight processes, and compared in Table~\ref{Par1}. The
greatest real-time speedup is obviously achieved by running with eight
processes, however, already using two processes helps significantly,
while the overhead (in terms of user time ratios) is very low.
\vspace{-3mm}
\begin{table*}[htbp]
  \caption{Comparison of the verification speed on article {\tt fdiff\_1} run in the normal mode and in the parallel mode, with eight parallel processes ({\tt -j8})}
\begin{center}
  \begin{tabular}{|l|r|r|r|r|}
    \hline
    Article&real (normal)&user (normal)&real (-j8)&user (-j8)\\
    \hline    
    fdiff\_1&13.11&12.99&3.54&21.20\\
    \hline    
  \end{tabular}
\end{center}
\label{Par}
\end{table*}
\vspace{-1.5cm}
%\vskip -1cm
% fdiff_1.miz.log_1:12.99user 0.01system 0:13.11elapsed 99%CPU (0avgtext+0avgdata 83008maxresident)k
% fdiff_1.miz.log_8:21.20user 0.44system 0:03.54elapsed 611%CPU (0avgtext+0avgdata 157760maxresident)k
% \begin{table*}[htbp]
%   \caption{Comparison of the verification speed on article {\tt fdiff\_2} run in the normal mode and in the parallel mode, with eight parallel processes ({\tt -j8})}
% \begin{center}
%   \begin{tabular}{|l|r|r|r|r|}
%     \hline
%     Article&real (normal)&user (normal)&real (-j8)&user (-j8)\\
%     \hline    
%     fdiff\_2&31.179&31.120&7.836&46.110\\
%     \hline    
%   \end{tabular}
% \end{center}
% \label{Par}
% \end{table*}
% time /home/mptp/gitrepo/MPTP2/MizAR/cgi-bin/bin/mizp.pl -j8    -x  ~/gitrepo/xsl4mizar -q -v /home/mptp/bin/verifier_7_11_04_parx  fdiff_2
% real    0m7.836s
% user    0m46.110s
% sys     0m1.190s
\begin{table*}[htbp]
  \caption{Comparison of the verification speeds on 395 slow MML articles run with one, two, four, and eight parallel processes}
\begin{center}
  \begin{tabular}{|l|r|r|r|r|}
    \hline
    & -j1 & -j2 & -j4 & -j8 \\
\hline
    Sum of user times (s)   & 12561.07 & 13289.41 & 15937.42 & 21697.71 \\
\hline
    Sum of real times (s)   & 13272.22 &  7667.37 & 5165.9 & 4277.12 \\
\hline
    Ratio of user time to -j1 & 1 & 1.06 & 1.27 & 1.73\\
\hline
    Ratio of real time to -j1  & 1 & 0.58 & 0.39 & 0.32 \\
    \hline    
  \end{tabular}
\end{center}
\label{Par1}
\end{table*}
\vspace{-1cm}
%12561.07,13272.22,424.06515312446,1.07630749523975
%13289.41,7667.37999999999,251.321030736207,0.637870636386312
%15937.42,5165.9,149.038829160942,0.37827114000239
%21697.71,4277.12,104.159660463366,0.264364620465396
When all the child instances finish their jobs, the parent
parallelizer postprocesses their results. In the case of running just
verification (Analyzer and Checker), the overall result is simply a
file containing the error messages and positions. This file is created
just by (uniquely) sorting together the error files produced by the
child instances.  Merging the HTML-ization results of the child
instances is very simple thanks to the mechanisms described in
Section~\ref{HTML}.  The {\tt --ajax-proofs} option is used to place
the HTMLized proofs into separate files, and depending on the required
HTML output, either just bound to AJAX calls in the toplevel
HTML-ization, inserting them on-demand, or postprocessing the toplevel
HTML in Perl by the direct inclusion of the HTML-ized toplevel proofs
into it (creating one big HTML file).

\subsection{Finer Parallelization}
The probably biggest practical disadvantage of the parallelization
based on toplevel proofs is that in some cases, the articles really
may consist of proofs with very uneven size, in extreme cases of just
one very large proof.  In such cases, the division of the toplevel
proofs into groups of similar size is going to fail, and the largest
chunk is going to take much more time in verification and HTML-ization
than the rest. One option is in such cases to recurse, and inspect the
sub-proof structure of the very long proofs, again, trying to
parallelize there.  This was not done yet, and instead, the
Checker-based parallelization was implemented, providing speedup just
for the most expensive Checker pass, but on the other hand, typically
providing a very large parallelization possibility.  This is now
implemented quite similarly to the toplevel proof parallelization, by
modifying the intermediate XML file passed from the Analyzer to the
Checker. As with the {\tt @proof} user-provided directive, there is a
similar internal directive usable in the XML file, telling the Checker
to skip the verification of a particular atomic inference. This is the
used very similarly to {\tt @proof}: The parallelizer divides the
atomic inferences into equally sized groups, and spawns $N$ children,
each of them modifying the intermediate XML file, and thus checking
only the inferences assigned to the particular child. The errors are
then again merged by the parent process, once all the child instances
have finished. 

The overall evaluation of this mode done again across all (395) MML
articles that take in the normal mode more than ten seconds to verify
is shown in Table~\ref{Par2} for (checker-only) -j8, and compared with the (toplevel) -j8 from
Table~\ref{Par1} where the toplevel parallelization mode is used. The
data confirm the general conjecture from Section~\ref{Types}: A lot of
Mizar's work is done in the type analysis module, and the opportunity to
parallelize that is missed in the Checker-only parallelization. This
results in lower overall user time (less work repetition in analysis), however higher real time (time perceived by
the user).
\vspace{-3mm}
\begin{table*}[htbp]
  \caption{Comparison of the toplevel and checker-only verification speeds on 395 slow MML articles run with one and eight parallel processes}
\begin{center}
  \begin{tabular}{|l|r|r|r|}
    \hline
    & -j1 & -j8 (toplevel) & -j8 (checker-only) \\
\hline
    Sum of user times (s)   & 12561.07 &  21697.71 & 18927.91 \\
\hline
    Sum of real times (s)   & 13272.22 &  4277.12 & 5664.1\\
\hline
    Ratio of user time to -j1 & 1 & 1.73 & 1.51 \\
\hline
    Ratio of real time to -j1  & 1 & 0.32 & 0.43 \\
    \hline    
  \end{tabular}
\end{center}
\label{Par2}
\end{table*}
\vspace{-8mm}
% 18927.91,5664.1,120.704113155185,0.30635561714514
This parallelization is in some sense orthogonal to the toplevel proof
parallelization, and it can be used to complement the toplevel proof
parallelization in cases when there are for instance only two major
toplevel proofs in the article, but the user wants to parallelize
more. I.e., it is no problem to recurse the parallelizer, using the
Checker-based parallelization for some of the child instances doing
toplevel-proof parallelization.

\section{Related Work}
As already mentioned, sophisticated parallelization and strategy
scheduling have been around in some ATP systems for several years now,
an advanced example is the infrastructure in the Waldmeister
system~\cite{Hillenbrand03}. The Large Theory Batch (LTB) division of
the CADE ATP System Competition has started to encourage such
development by allowing parallelization on multicore competition
machines. This development suits particularly well the ATP/LTB tasks
generated in proof assistance mode for Mizar. Recent parallelization
of the Isabelle proof assistant and its implementation language are
reported in~\cite{MatthewsW10} and in~\cite{WenzelM09}, focusing on
fitting parallelism within the LCF approach. This probably makes the
setting quite different: \cite{WenzelM09} states that \textit{there is
  no magical way to add the “aspect of parallelism” automatically},
which does not seem to be the case with the relatively straightforward
approaches suggested and used here for multiple parts of Mizar and
related processing. As always, there seems to be a trade-off between
(in this case LCF-like) safety aspirations, and efficiency, usability,
and implementation concerns. Advanced ITP systems are today much more
than just simple slow proof checkers, facing similar ``safety''
vs. ``efficiency'' issues as ATP systems~\cite{ivy}.  The Mizar
philosophy favors (sometimes perhaps too much) the latter, arguing
that there are always enough ways how to increase certainty, for
example, by cross-verification as in~\cite{Urban-GDV}, which has been
recently suggested as a useful check even for the currently safest
LCF-like system in~\cite{Adams10}. Needless to say, in the particular
case of parallelization a possible error in the parallelization code
is hardly an issue for any proof assistant (LCF or not) focused on
building large libraries. As already mentioned in
Section~\ref{MizarProcessing}, at least in case of Mizar the whole
library is typically re-factored and re-verified many times, for which
the safe file-based parallelization is superior to internal
parallelization also in terms of efficiency, and this effectively
serves as overredundant automated cross-verification of the internal parallelization code.

% While having safety is always good, a possible error
% in the parallelization code is hardly an issue for any proof assistant
% (LCF or not), as the whole library is typically re-verified many
% times, for which the safe file-based parallelization is superior 
% to internal parallelization also
% in terms of efficiency.

\vspace{-1.2mm}
\section{Future Work and Conclusions}
\label{Future}
The parallelizer has been integrated in the Mizar mode for
Emacs~\cite{Urban06-jal} and can be used instead of the standard
verification process, provided that Perl and LibXML are installed, and
also in the remote server verification mode, provided Internet is
available. The speedups resulting from combination of these two
techniques are very significant. As mentioned above, other Mizar
utilities than just the standard verifier can be parallelized in
exactly the same way, and the Emacs environment allows this too. 
% Work
% in this direction seems to be useful both for providing the fast
% server-based verification, but also for the users working on their
% desktops and notebooks.  
The solutions described in this paper might
be quite Mizar-specific, and possibly hard to port e.g., to systems
with non-opaque proofs like Coq, and the LCF-based provers, that do
not use similar technique of compilation-like passes. Other, more
mathematician-oriented Mizar-like systems consisting of separate
linguistic passes like SAD/ForThel~\cite{LyaletskiV10} and
Naproche~\cite{CramerFKKSV09} might be able to re-use this approach
more easily.
% Multithreading solutions (like
% MTProcs for Free Pascal) seem to exist for many other languages, and
% actually the inspiration for this work came from discussion with
% Makarius Wenzel about the usage of multithreading in PolyML for
% parallelizing Isabelle.  One of the advantages of the multithreading solutions is
% that once the refactoring of the code is done, the implementation can
% typically be compiled on many platforms, and does not require further
% software (like Perl, Unix (for doing {\tt fork()}), and LibXML) like
% in the work described here.  This is actually quite important for
% Mizar, with its relatively high number of MS Windows users, so an
% MTProcs-based multithreading re-implementation of the Checker would be
% interesting from this point of view.

As mentioned above, another motivation for this work comes from the
work on a wiki for formal mathematics, and for that mode of work it
would be good to have finer dependencies between the the various items
introduced and proved in the articles. Once that is available, the
methods developed here for file-based parallelization will be also
usable in a similar way for minimalistic checking of only the selected
parts of the articles that have to be quickly re-checked due to some
change in their dependencies. This %``finer dependencies'' 
mode of work
thus seems to be useful to have not just for Mizar, but for any %formal
proof assistant that would like to have its library available,
editable, and real-time verifiable in an online web repository.

\bibliographystyle{plain}
\begin{small}
\bibliography{ate14,bib1}
\end{small}
\end{document}